\documentclass{iaus}
\usepackage{graphicx}
\usepackage{times}

\title[Class I methanol masers] 
{The Australia Telescope campaign to study southern class I methanol masers}

\author[Voronkov et al.]   
{M.A.Voronkov$^1$,  
 K.J.Brooks$^1$, A.M.Sobolev$^2$, S.P.Ellingsen$^3$, A.B.Ostrovskii$^2$,
 J.L.Caswell$^1$}

\affiliation{$^1$Australia Telescope National Facility, PO Box 76,
Epping, NSW 1710 \break email: Maxim.Voronkov@csiro.au\\[\affilskip]
$^2$Ural State University, Lenin ave. 51, 620083 Ekaterinburg, Russia\\[\affilskip]
$^3$University of Tasmania, GPO Box 252-37, Hobart, Tasmania 7000, Australia}

\pubyear{2007}
\volume{242}  
\pagerange{1--2}
\date{?? and in revised form ??}
\jname{Proceedings Title IAU Symposium}
\editors{J. Chapman \& W.Baan, eds.}

\begin{document}

\maketitle

\begin{abstract}
The Australia Telescope Compact Array (ATCA) and the Mopra facility have
been used to search for new southern class I methanol masers at
9.9, 25 (J=5) and 104 GHz, which are thought to trace more energetic
conditions in the interface regions of molecular outflows, than 
the widespread class I masers at 44 and 95 GHz. One source shows a clear
outflow association.
\end{abstract}


\begin{figure}
 \includegraphics[width=\linewidth]{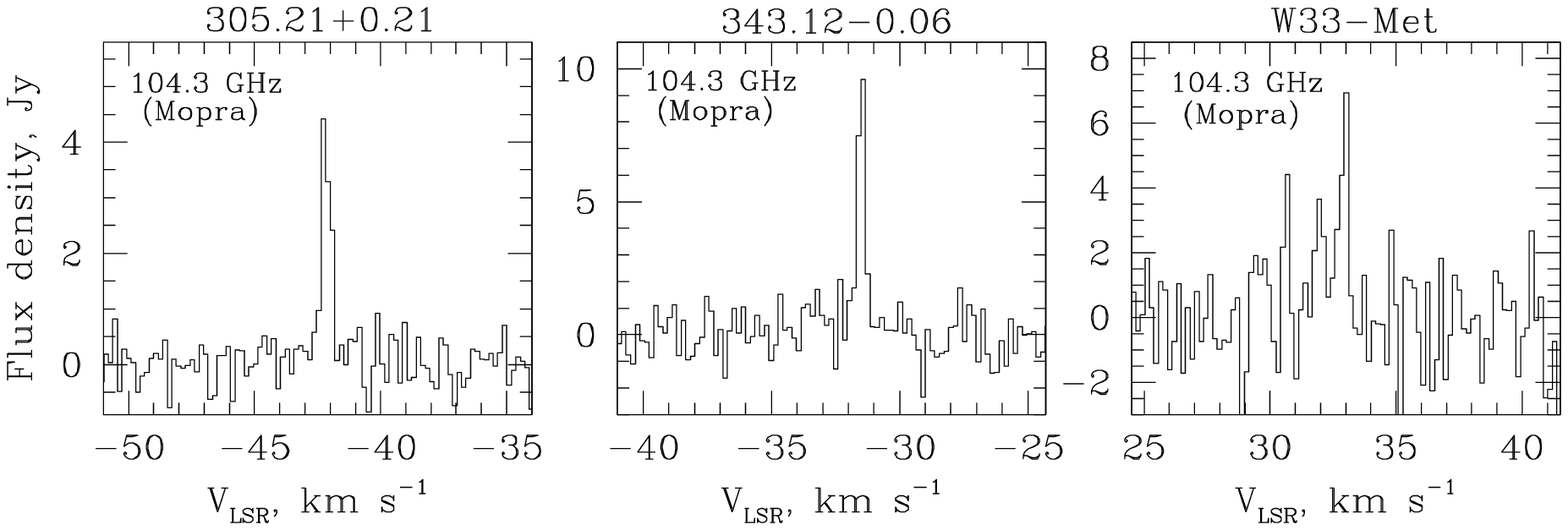}
 \vskip -5mm
 \includegraphics[width=\linewidth]{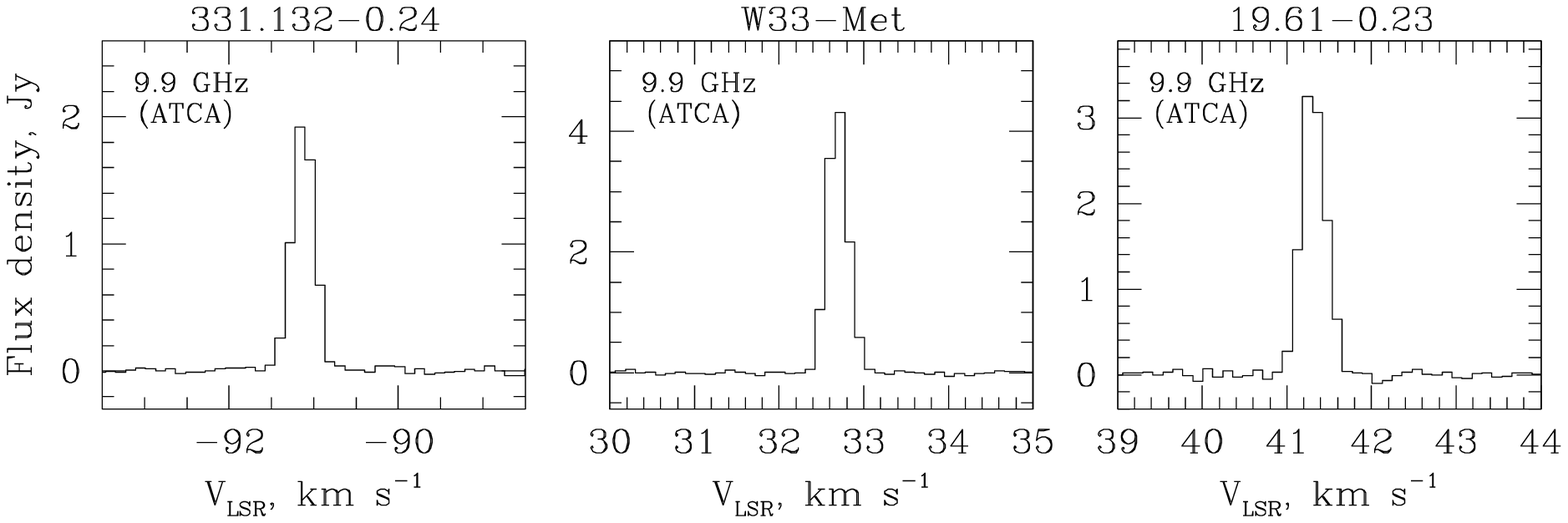}
 \vskip -3mm
 \includegraphics[width=\linewidth]{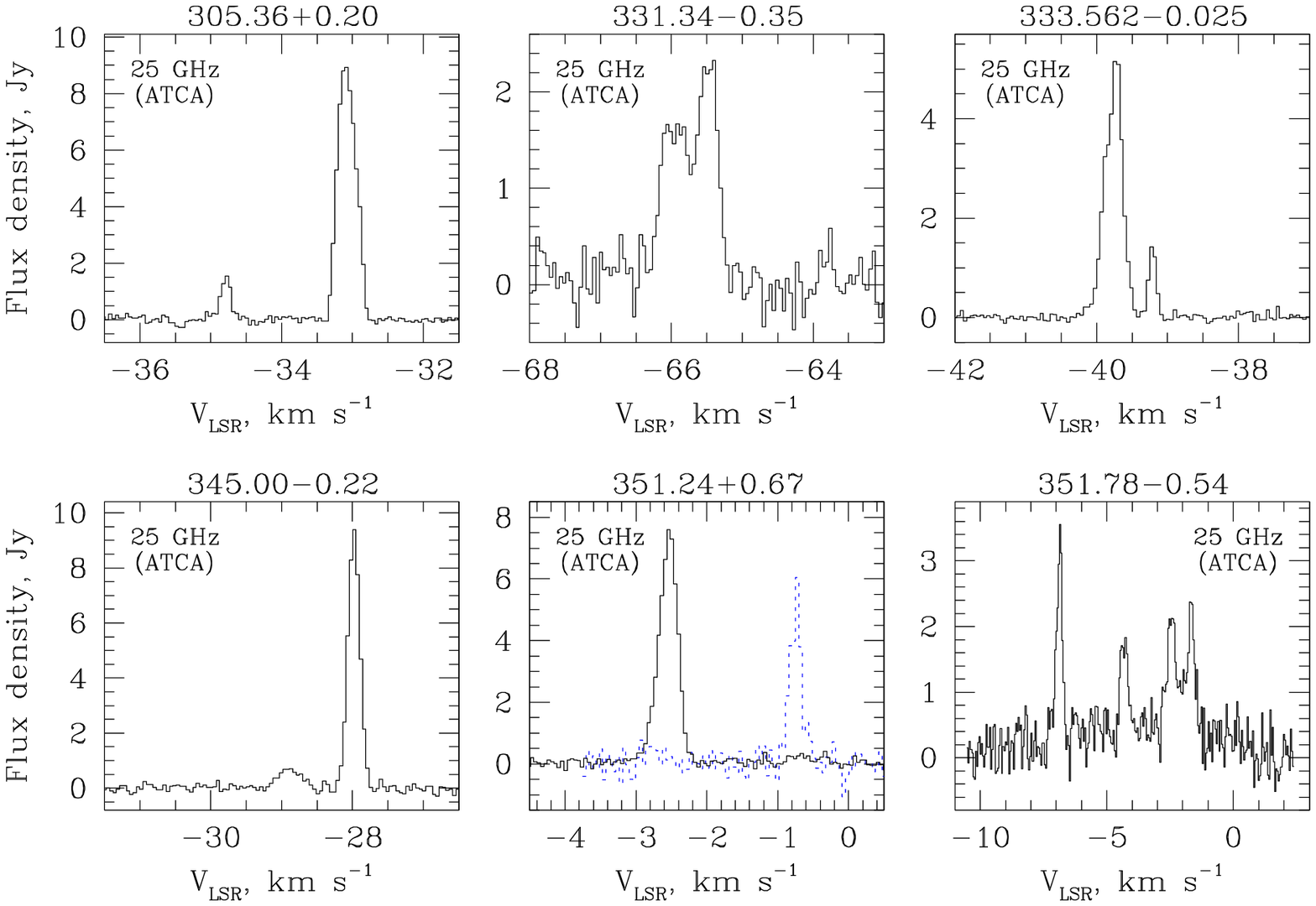}
 \caption{The spectra of the 104 GHz (top row), 9.9 GHz (second row) and
 25 GHz (two bottom rows) maser detections. The 25~GHz spectra are given
 only for the sources with measured accurate positions. The
 dotted (blue) spectrum for G351.24$+$0.67 corresponds to an offset
 component (see the poster referred to in the text).}\label{fig:spectra}
\end{figure}

A number of surveys have been carried out to investigate various class I
methanol masers and to search for an association with the other constituents of
a typical massive star-forming region. The ATCA 9.9 and Mopra 104 GHz
surveys indicate that these masers are rare (3 detections out
of 40 targets and 2 detections out of 69 targets, respectively) confirming
the results of the previous searches
\cite[(Slysh et al., 1993; Voronkov et al., 2005)]{vor05,sly93}.
In contrast, there were 67 detections at 25 GHz (J=5) out of 102 targets
observed (ATCA observations, but no accurate positions were measured for
the majority of the sources), although these masers were also previously
believed to be rare following a survey of \cite{men86}. Unlike the spectra of the 44 GHz masers, the
spectra of 9.9, 25 and 104 GHz masers appeared to be quite simple (small
number of components) and no significant correlation between the flux
densities has been found. It is in line with the theoretical
prediction that these masers require more energetic conditions
(higher temperatures and densities) to form \cite[(Sobolev et al., 2005)]{sob05}.
Only one source, G343.12-0.06 (IRAS 16547-4247), was active in all observed
transitions. Its detailed observations showed a remarkable association
with a jet-driven molecular outflow (see \cite[Voronkov et al., 2006]{vor06}
for details). The spectra of detections are given in Fig.\ref{fig:spectra}.
The 25 GHz spectra are given only for the sources with measured accurate
positions. A table with the absolute positions of maser spots is available
in the poster, which can be downloaded from
{\small\it http://www.atnf.csiro.au/research/masermeeting/web\_papers/333\_Voronkov.pdf}

\end{document}